\documentclass[12pt]{iopart}

\usepackage{graphicx}
\usepackage{subfigure}
\usepackage{verbatim}
\usepackage{dcolumn}
\usepackage{bm}
\usepackage{epsf}
\usepackage{color}
\usepackage[toc]{appendix}
\usepackage{stmaryrd}
\usepackage{amsthm}
\usepackage{hhline}
\usepackage{amssymb}
\usepackage{iopams}
\usepackage{cite}
\usepackage{tikz}
\usetikzlibrary{arrows,shapes,calc,matrix}
\expandafter\let\csname equation*\endcsname\relax
\expandafter\let\csname endequation*\endcsname\relax
\usepackage{amsmath}
\usepackage{xstring}
\usepackage[colorlinks=true,citecolor=blue,linkcolor=blue,urlcolor=blue]{hyperref}

\begin{document}


\title{Boosting engine performance with Bose-Einstein condensation}

\author{Nathan M. Myers\textsuperscript{1,2,7}, Francisco J. Pe\~na\textsuperscript{3,7}, Oscar Negrete\textsuperscript{4}, Patricio Vargas\textsuperscript{4}, Gabriele De Chiara\textsuperscript{5} and Sebastian Deffner\textsuperscript{1,6}}
\address{$^1$ Department of Physics, University of Maryland, Baltimore County, Baltimore, Maryland 21250, USA}
\address{$^2$Computer, Computational and Statistical Sciences Division, Los Alamos National Laboratory, Los Alamos, New Mexico 87545, USA}
\address{$^3$ Departamento de Física, Universidad Técnica Federico Santa María, Casilla 110V, Valparaíso, Chile}
\address{$^4$ Departamento de Física, CEDENNA, Universidad Técnica Federico Santa María, Casilla 110V, Valparaíso, Chile}
\address{$^5$Centre for Theoretical Atomic, Molecular and Optical Physics,
Queen’s University Belfast, Belfast BT7 1NN, United Kingdom}
\address{$^6$Instituto de F\'{i}sica `Gleb Wataghin', Universidade Estadual de Campinas, 13083-859, Campinas, S\~{a}o Paulo, Brazil}
\address{$^7$Authors to whom any correspondence should be addressed.}
\ead{myersn1@umbc.edu (N.M.M.), francisco.penar@usm.cl (F.J.P.)}

\begin{abstract}
At low-temperatures a gas of bosons will undergo a phase transition into a quantum state of matter known as a Bose-Einstein condensate (BEC), in which a large fraction of the particles will occupy the ground state simultaneously. Here we explore the performance of an endoreversible Otto cycle operating with a harmonically confined Bose gas as the working medium. We analyze the engine operation in three regimes, with the working medium in the BEC phase, in the gas phase, and driven across the BEC transition during each cycle. We find that the unique properties of the BEC phase allow for enhanced engine performance, including increased power output and higher efficiency at maximum power.
\end{abstract}

\section{Introduction}
 
 In the 1920s Bose \cite{bose1924plancks} and Einstein \cite{einstein1924sitzber} put forward the theoretical hypothesis that a dilute atomic gas could give way to a phenomenon in which a large number of bosons occupy the zero momentum state of a system simultaneously. This phenomenon, now known as Bose-Einstein condensation (BEC), was corroborated in 1995 when it was observed in rubidium \cite{anderson1995observation}, sodium  \cite{davis1995bose} and lithium \cite{bradley1995evidence, bradley1997bose} vapors, confined in magnetic traps and cooled to temperatures in the fractions of microkelvins in order to achieve the necessary ground state populations. These experimental verifications marked a profound development in the study of quantum gases. Over the subsequent years, experimental control of BECs has expanded dramatically, including the creation of a BEC in microgravity \cite{Becker2018} and the implementation of BEC-based atomic circuits \cite{Ramanathan2011}.

As a phase transition with an origin that is purely quantum in nature, the thermodynamics of BECs has attracted considerable attention. The transition from a normal Bose gas to a BEC can be fully described mathematically, and treatments can be found in most any modern thermodynamics or statistical mechanics textbook \cite{Huang2009book, Pathria2011, Pitaevskii2016book}. Notably, unlike the more familiar gas-to-liquid phase transition, the BEC transition occurs in momentum, rather than coordinate space \cite{Pathria2011}. While the equilibrium thermodynamic behavior of BECs is well established, including the equations of state, fugacity, and specific heat \cite{Huang2009book, Pathria2011, Pitaevskii2016book}, the analysis of BECs in the context of heat engines, the paradigmatic systems that thermodynamics itself was developed to study, remains curiously scarce.    
 
With the development of quantum thermodynamics \cite{Deffner2019book} the exploration of how quantum phenomena can be harnessed in nanoscale thermal machines has seen an explosion in interest \cite{Scully2003, Scully2011, Abah2014, Rossnagel2014, Hardal2015, Manzano2016, Niedenzu2016, Watanabe2017, Klaers2017, Friedenberger2017, Deffner2018}. With macroscopically observable quantum features and well-developed techniques for experimental control, Bose-Einstein condensates would seem an optimal system to serve as a quantum working medium for a thermal machine. However, as the condensate itself consists of macroscopic occupation of the zero-momentum state, it is not easy to see how the typical paradigm for work extraction from macroscopic thermal machines, involving pressure exerted against an external piston or potential, translates to a BEC working medium. Several recent works have proposed implementations of quantum thermal machines that leverage BECs, including extracting work through the use of Feshbach resonances \cite{Li2018}, using a mixture of two gas species to implement a refrigeration cycle \cite{Niedenzu2019}, implementing a heat engine cycle with cold bosons confined to a double-well potential \cite{Fialko2012}, and using BECs as the basis for thermal machines that act on a working medium of quantum fields \cite{Gluza2021}.

In this paper, we explore an Otto cycle in the context of endoreversible thermodynamics using a harmonically trapped bosonic gas as a working substance. We study the cycle in three regions of operation: i) with a condensed medium, ii) with a non-condensed medium, and iii) with a medium driven across the condensation transition. We find that the properties of the BEC allow for enhanced performance above what can be achieved with the corresponding classical gas. For a working medium that remains in the condensate phase during the whole cycle, we show that the efficiency at maximum power significantly exceeds the Curzon-Ahlborn (CA) efficiency \cite{Curzon1975}, the efficiency obtained for an endoreversible Otto cycle with a working medium of an ideal gas described by Boltzmann statistics \cite{Leff1987}. In contrast, if the system only operates with the working medium in a non-condensed phase, we find that the efficiency at maximum power is equivalent to the CA efficiency. We also examine cycles operating while the working medium is driven across the BEC phase transition, and find that the efficiency at maximum power is highly parameter-dependent and can fall above or below the CA efficiency. We conclude with a discussion on the role of the condensate itself in work extraction and the experimental applicability of these results.  
 
\section{BEC thermodynamics}
 
To keep our analysis as self-contained as possible and establish the necessary notions and notations, we begin with a brief review of the textbook thermodynamics of non-interacting bosons in a harmonic trap. We consider the system under study to be in the thermodynamic limit, in which $N\rightarrow \infty$, where $N$ is the number of bosons, while maintaining the condition $N\omega$ is constant, where $\omega$ corresponds to the trap frequency \cite{Pitaevskii2016book}. The trapping potential is given by, 
\begin{equation}
\label{potential}
    V_{\mathrm{ext}}(\mathbf{r})=\frac{1}{2}m\left(\omega_{x}^{2}x^{2} + \omega_{y}^{2}y^{2} + \omega_{z}^{2}z^{2} \right),
\end{equation}
 where $\omega_{x}$, $\omega_{y}$ and $\omega_{x}$ are the oscillator frequencies in each direction. The energy eigenvalues for each atom of the Hamiltonian corresponding to the above potential are \cite{Pathria2011, Pitaevskii2016book},
 \begin{equation}
 \label{energyspectrum}
      E_{n_{x},n_{y},n_{z}}=\hbar \omega_{x} \left(n_{x} + \frac{1}{2}\right) + \hbar \omega_{y} \left(n_{y}+\frac{1}{2}\right) + \hbar \omega_{z} \left(n_{z}+\frac{1}{2}\right). 
 \end{equation}
If all three frequencies are the same (the harmonic-isotropic case), we can define $n=n_{x}+n_{y}+n_{z}$, simplifying the energy spectrum of Eq.~(\ref{energyspectrum}) to $E_{n}=\hbar \omega \left(n+\frac{3}{2}\right)$ with a quantum degeneracy of the form $\mathcal{D}(n)=\left(n + 1\right) \left(n + 2\right)/2$ \cite{Pathria2011}.

In the harmonic-isotropic case, the Grand Potential for a system of bosons in the Grand Canonical ensemble is given by \cite{Pathria2011},
\begin{equation}
    \Omega (\mu, T, \omega)=k_{B} T \sum_{n_{x},n_{y}, n_{z}} \ln\left(1-e^{-\beta \hbar \omega \left(n_{x} + n_{y} + n_{z}\right) + \beta\mu}\right),
\end{equation}
where we have suppressed the zero energy state in order to obtain the number of excited bosons in the system. We can perform the above sum by introducing a continuous density of states (assuming that $E \gg \hbar\omega$) \cite{Pathria2011, Pitaevskii2016book},
\begin{equation}
\label{densityofstates}
    a(E)=\frac{E^{2}}{2 \left(\hbar \omega\right)^{3}}.
\end{equation}
In this approximation, the Grand Potential takes the form \cite{Pathria2011},
\begin{equation}
\label{grandpotential}
     \Omega(\mu, T, \omega)=\frac{(k_{B} T)^{4}}{2(\hbar \omega)^{3}}\int^{\infty}_{0}x^{2}\ln\left(1-e^{-x}e^{\beta\mu}\right)=-\frac{(k_{B} T)^{4}}{(\hbar \omega)^{3}}g_{4}(z),
 \end{equation}
where $\beta$ is the inverse temperature, $\mu$ is the chemical potential, and $g_{4}(z)$ corresponds to the Bose function, given by the integral \cite{Pathria2011, Pitaevskii2016book}, 
\begin{equation}
\label{bosefunction}
    g_{\nu}(z)=\frac{1}{\Gamma(\nu)}\int_{0}^{\infty} dx \frac{x^{\nu-1}}{z^{-1}e^{x}-1},
\end{equation}
or in series form as \cite{Pathria2011, Pitaevskii2016book},
\begin{equation}
\label{eq:bosefunctionsum}
    g_{\nu}(z) = \sum_{n = 1}^{\infty} \frac{z^n}{n^{\nu}}.
\end{equation}
Here $z=\mathrm{exp}(\mu/k_{B}T)$ denotes the fugacity of the system. Note that, for the case of harmonic confinement, the volume is not a parameter in the Grand Potential. Instead, the inverse of the trap frequency plays the role of volume. The average number of excited atoms in the trap can be obtained from \cite{Pathria2011, Pitaevskii2016book},
\begin{equation}
    \label{nthermodynamic}
    N(\mu,T)=-\left(\frac{\partial \Omega}{\partial \mu}\right)_{\omega, T}=\left(\frac{k_{B} T}{\hbar \omega}\right)^{3}g_{3}(z),
\end{equation}
where we use the recurrence relation \cite{Pathria2011},
\begin{equation}
g_{\nu - 1} (z)=\frac{\partial}{\partial \ln(z)} g_{\nu}(z).
\end{equation}

For fixed $N$, the fugacity monotonically increases as temperature decreases until Bose-Einstein condensation occurs at $\mu=0$ $(z=1)$ \cite{Pathria2011}. Therefore, using Eq.~(\ref{nthermodynamic}) and setting $z=1$, we can find the critical temperature that characterizes the transition \cite{Pathria2011, Pitaevskii2016book}, 
\begin{equation}
\label{critical}
    T_{c}=\frac{\hbar \omega}{k_{B}}\left(\frac{N}{\zeta(3)}\right)^{\frac{1}{3}}.
\end{equation}
 
The internal energy of the system below and above the BEC transition is given by \cite{Pathria2011, Pitaevskii2016book},
\begin{eqnarray}
 \label{internalenergy}
  U(T, \omega)=\left\{
    \begin{array}{ll}
      3 k_{B}T\left(\frac{k_{B}T}{\hbar \omega}\right)^{3} g_{4}(1), & \mbox{$T \leq T_{c}$}.\\
      3 k_{B}T\left(\frac{k_{B}T}{\hbar \omega}\right)^{3} g_{4}(z), & \mbox{$T \geq T_{c}$}.
    \end{array}
  \right.
\end{eqnarray}
Using Eq. \eqref{internalenergy} the entropy of the system can be found \cite{Pitaevskii2016book}, 
\begin{eqnarray}
 \label{entropy}
  S(T, \omega)=\left\{
    \begin{array}{ll}
      4 k_{B}\left(\frac{k_{B}T}{\hbar \omega}\right)^{3} g_{4}(1), & \mbox{$T \leq T_{c}$}.\\
       k_{B}\left(\frac{k_{B} T}{\hbar \omega}\right)^{3}\zeta (3) \left[\frac{4}{N}\left(\frac{k_{B} T}{\hbar \omega}\right)^{3}g_{4}(z)-\ln (z)\right], & \mbox{$T \geq T_{c}$}.
    \end{array}
  \right.
\end{eqnarray}
Note that when using Eq.~(\ref{entropy}) we can obtain an expression for the fugacity for the case of $T \geq T_{c}$ by solving Eq.~(\ref{nthermodynamic}).


\section{The endoreversible Otto cycle}

\begin{figure}[!ht]  
\center
\includegraphics[width=0.8 \textwidth]{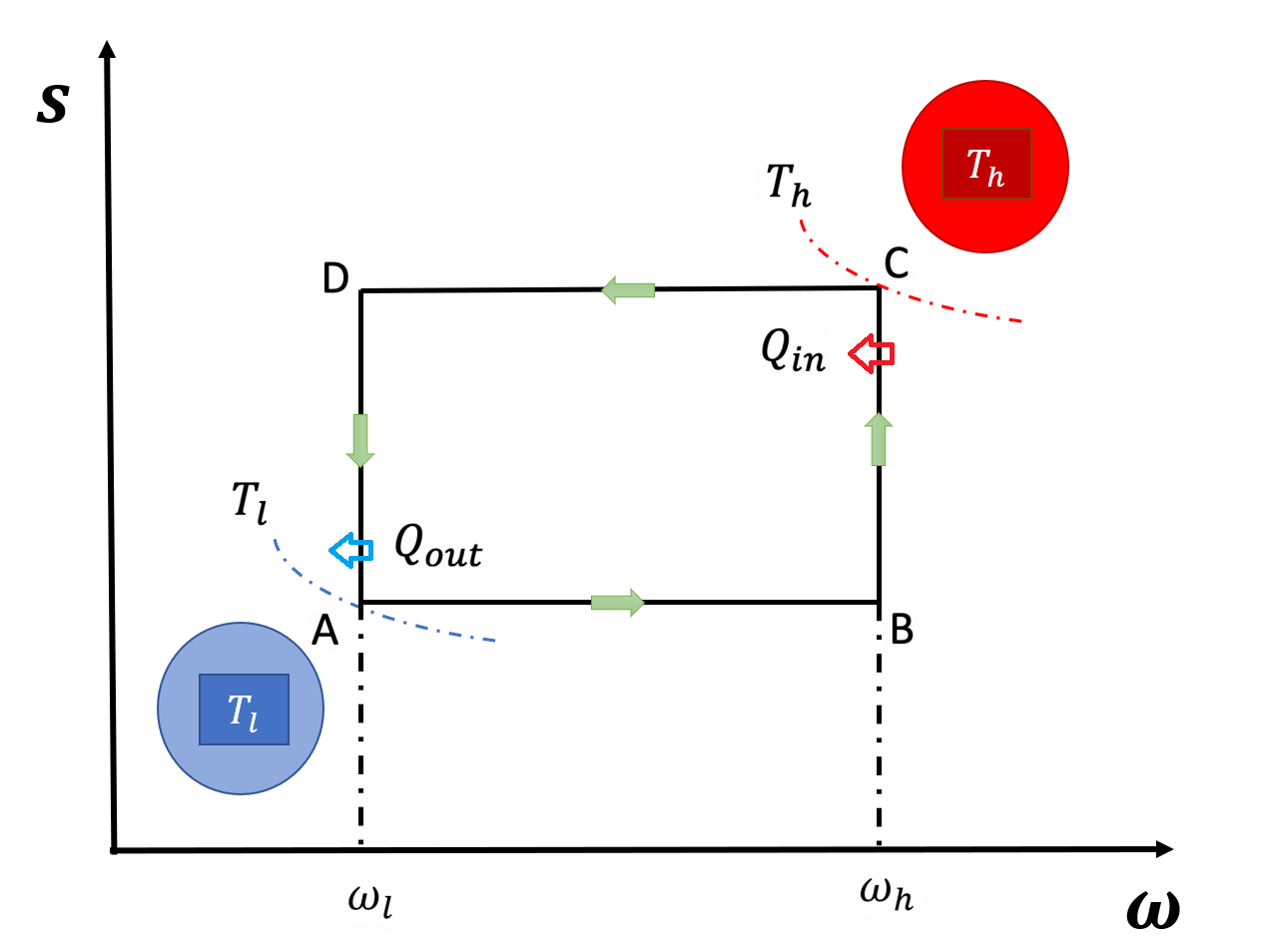}
 \caption{Entropy versus external field diagram for the Otto Cycle. Note that the system is only in contact with the thermal reservoirs during the isochoric (vertical) strokes. At points $\mathrm{C}$ and $\mathrm{A}$, the working substance reaches the temperatures $T_{h}$ and $T_{l}$, of the hot and cold reservoirs, respectively, indicated by the isotherms touching the cycle at those points. For the quantum cycle, the entropy values $S_{\mathrm{B}}$ and $S_{\mathrm{D}}$  are calculated using the same thermal occupation probabilities as points $\mathrm{A}$ and $\mathrm{C}$ to ensure the strokes $\mathrm{A} \rightarrow\mathrm{B}$ and $\mathrm{C} \rightarrow\mathrm{D}$ fulfill the quantum adiabatic condition.}
 \label{OttoCycle}
\end{figure}

The Otto cycle consists of four strokes: isentropic compression, isochoric heating, isentropic expansion, and isochoric cooling. The cycle strokes for a working medium of harmonically confined particles are illustrated graphically in Fig. \ref{OttoCycle} using an entropy ($S$) - frequency ($\omega$) diagram. The isentropic and isochoric processes are represented in the figure by horizontal and vertical lines, respectively. In our notation, $T$ refers to temperature and $\omega$ to the trap frequency (both parameters measured in arbitrary units). During the isentropic strokes the working system is disconnected from the thermal reservoirs and the external field is varied from $\omega_{l}$ to $\omega_{h}$ (for stroke $\mathrm{A}\rightarrow \mathrm{B}$) and vice-versa (for stroke $\mathrm{C}\rightarrow \mathrm{D}$). In contrast, during the isochoric strokes the external field is held constant while the working medium exchanges heat with the hot (for stroke $\mathrm{B}\rightarrow \mathrm{C}$) or cold (for stroke $\mathrm{D}\rightarrow \mathrm{A}$) reservoir. Note that the work parameter ($\omega$) plays the role of an \textit{inverse} volume, increasing during the compression stroke ($\mathrm{A} \rightarrow \mathrm{B}$) and decreasing during the expansion stroke ($\mathrm{C} \rightarrow \mathrm{D}$).  

Thermodynamically, the cycle is characterized by the temperatures of the two heat reservoirs and the initial and final values of the external frequency, $\omega_{h}$ and $\omega_{l}$. The finite-time performance of the cycle can be analyzed using the framework of \textit{endoreversible thermodynamics} \cite{Curzon1975, Rubin1979, Hoffmann1997}. Note that the finite-time analysis of an Otto cycle with a working medium of an ideal Bose gas was previously examined in Ref. \cite{Wang2009}. However, in that study the effects of Bose-Einstein condensation were not explored, which we will examine in detail. For our analysis we will closely follow the procedure established in Ref. \cite{Deffner2018}. During an endoreversible process the working medium is assumed to always be in a state of local equilibrium, but never achieves global equilibrium with the reservoirs. As such, we can express the heat exchanged with the reservoirs during the isochoric heating stroke (from $\mathrm{B} \rightarrow \mathrm{C}$) as, 
\begin{equation}
\label{qinendo}
Q_{\mathrm{in}}=U_{\mathrm{C}}(T_{3},\omega_{h})-U_{\mathrm{B}}(T_{\mathrm{2}},\omega_{h}),
\end{equation}
where we note that, unlike the quasistatic case, $T_3 \neq T_h$. The temperatures $T_{2}$ and $T_{3}$ satisfy the following conditions,
\begin{equation}
    T(0)= T_2 \quad \mathrm{and} \quad T(\tau_{h})= T_{3} \quad \mathrm{with} \quad T_2 < T_{3} \leq T_{h},
\end{equation}
where $\tau_{h}$ is the duration of the heating stroke. We can explicitly model the temperature change from $T_2$ to $T_{3}$ by applying Fourier's law of heat conduction,
\begin{equation}
\label{isochoricheatingdiff}
    \frac{d T}{d t} =-\alpha_{h}\left(T(t)- T_{h}\right),
\end{equation}
where $\alpha_{h}$ is a constant that depends on the thermal conductivity and heat capacity of the working medium. Equation (\ref{isochoricheatingdiff}) can be fully solved to yield, 
\begin{equation}
\label{eq:isoheating}
    T_{3} - T_{h}= (T_2 - T_{h}) e^{-\alpha_{h}\tau_{h}}.
\end{equation}

The isentropic expansion stroke (from $\mathrm{C} \rightarrow \mathrm{D}$) is carried out in exactly the same manner as in the quasistatic cycle. Since the working medium is decoupled from the thermal reservoirs during this stroke, the work is determined entirely from the change in internal energy,
\begin{equation}
	\label{eq:Wexp}
	W_{\mathrm{exp}} = U_{\mathrm{D}}(T_4, \omega_l) - U_{\mathrm{C}}(T_3, \omega_h).
\end{equation} 

The isochoric cooling stroke (from $\mathrm{D} \rightarrow \mathrm{A}$) can be modeled in the exact same manner as the heating stroke. The heat exchanged with the cold reservoir is given by,
\begin{equation}
\label{qoutendo}
Q_{\mathrm{out}}=U_{\mathrm{A}}(T_{1},\omega_{l})-U_{\mathrm{D}}(T_4,\omega_{l}),
\end{equation}
where $T_{1}$ and $T_4$ satisfy the conditions
\begin{equation}
    T(0)= T_{\mathrm{D}} \quad \mathrm{and} \quad T(\tau_{l})= T_{1} \quad \mathrm{with} \quad T_4 > T_{1} \geq T_{l}.
\end{equation}
As for the heating stroke, the temperature change can again be modeled by Fourier's law, 
\begin{equation}
\label{isochoriccoolingdiff}
    \frac{d T}{d t} =-\alpha_{l}\left(T(t)- T_{l}\right),
\end{equation}
The solution to Eq.~(\ref{isochoriccoolingdiff}) is,
\begin{equation}
\label{eq:isocooling}
    T_{1} - T_{l} =\left(T_4 - T_{l}\right)e^{-\alpha_{l}\tau_{l}}.
\end{equation}

Finally, in exact analogy to the expansion stroke, the work done during the compression stroke can be found from the change in internal energy,
\begin{equation}
	\label{eq:Wcomp}
	W_{\mathrm{comp}} = U_{\mathrm{B}}(T_2, \omega_h) - U_{\mathrm{A}}(T_1, \omega_l).
\end{equation} 

The efficiency of the engine can be found from the ratio of the total work and the heat exchanged with the hot reservoir,
\begin{equation}
	\label{eq:eff}
	\eta = -\frac{W_{\mathrm{comp}}+W_{\mathrm{exp}}}{Q_{\mathrm{in}}}.
\end{equation}
The power output is given by the ratio of the total work to the cycle duration,
\begin{equation}
	\label{eq:P}
	P = -\frac{W_{\mathrm{comp}}+W_{\mathrm{exp}}}{\gamma (\tau_h + \tau_l)},
\end{equation}
with $\gamma$ serving as a multiplicative factor that implicitly incorporates the duration of the isentropic strokes \cite{Deffner2018}.

Noting that the entropy remains constant during the isentropic strokes, we can solve the equation for the total entropy differential $dS(T,\omega)=0$ to obtain a relationship between $T$ and $\omega$. This first order differential equation is given by,
\begin{equation}
\label{differential}
    \frac{d\omega}{dT}=-\frac{\left(\frac{\partial S}{\partial T}\right)_{\omega}}{\left(\frac{\partial S}{\partial \omega}\right)_{T}}.
\end{equation}
Taking the appropriate derivatives and solving Eq. \eqref{differential} we find that the isentropic condition is satisfied by,
\begin{equation}
\label{temperaturerelationadiabatic}
T_{2} = T_{1}\left(\frac{\omega_{h}}{\omega_{l}}\right) \equiv T_{1} \kappa^{-1}, \quad  T_{4} = T_{3}\left(\frac{\omega_{l}}{\omega_{h}}\right) \equiv T_{3} \kappa.
\end{equation}
Where $\kappa = \omega_{1}/\omega_{2}$ indicates the compression ratio. The full derivation of this condition is given in Appendix \hyperref[sec:AppendixA]{A}.

We note that the quasistatic Otto cycle can be recovered from the endoreversible cycle in the limit of long stroke times $\tau_h$ and $\tau_l$. In this limit, the working medium reaches full equilibrium with the reservoirs during the heating and cooling strokes and Eqs. \eqref{eq:isoheating} and \eqref{eq:isocooling} simplify to $T_3 = T_h$ and $T_1 = T_l$, respectively.

\section{Results}

\subsection{Quasistatic results}

\begin{figure}
	\centering
	\includegraphics[width=1\textwidth]{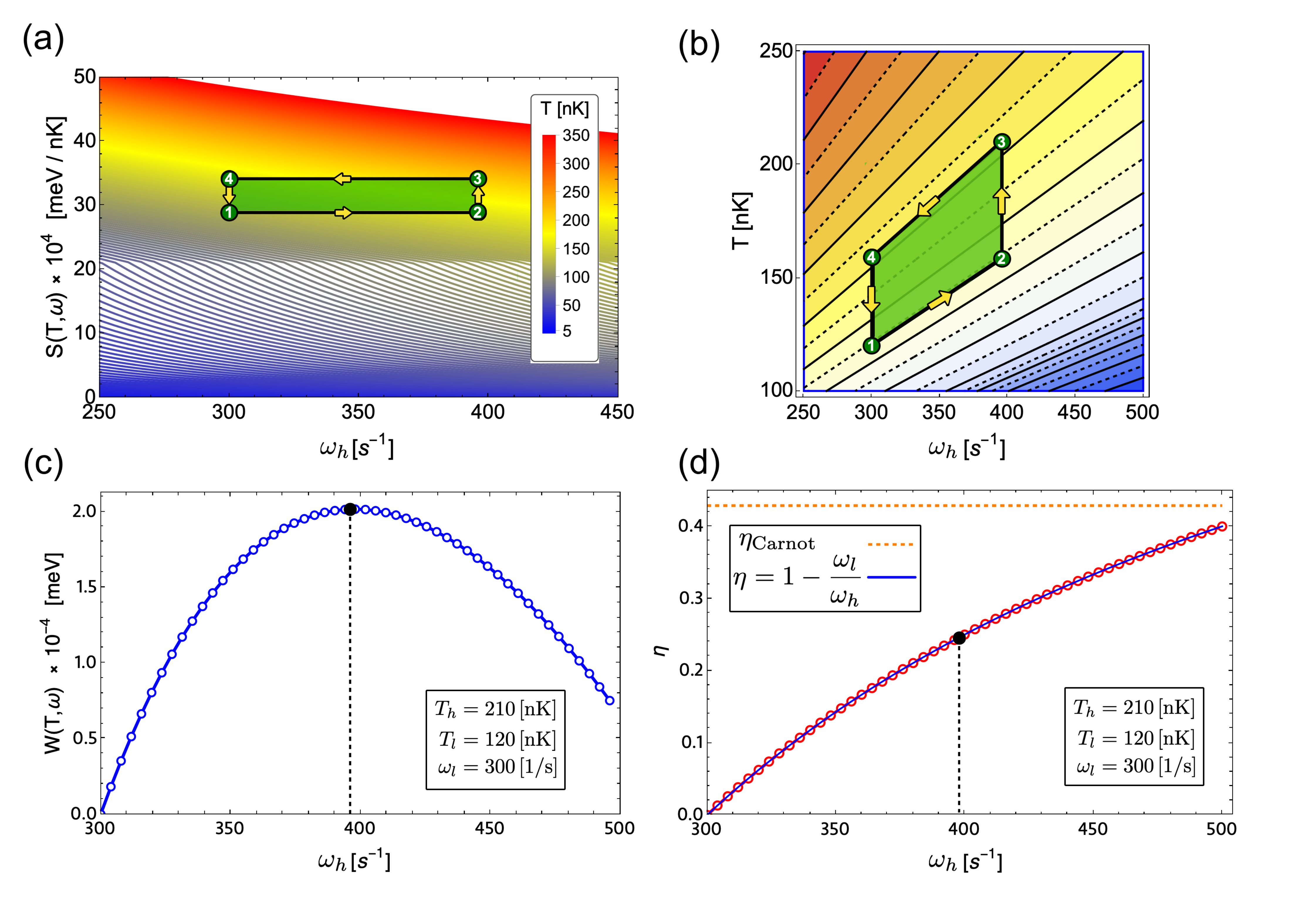}
	\caption{\label{fig:noncondensate} Results for a quasistatic cycle with 60,000 bosons operating fully in the non-condensate regime. (a) Representative cycle displayed over isothermal entropy curves. (b) $T$ versus $\omega_h$ plot of the cycle depicted in panel (a). (c) Total work as a function of $\omega_h$ for a range of cycles. (d) Efficiency as a function of $\omega_h$ for a range of cycles, with the analytical efficiency (solid blue line) and Carnot efficiency (dotted orange line) given for comparison. In panels (c) and (d) the work and efficiency values indicated with the black dots correspond to the specific cycle drawn in (a) and (b). The parameters are: $T_{l}= 120$ nK, $T_{h}=210$ nK, and $\omega_{l}=300 s^{-1}$.}
\end{figure}

\begin{figure}
	\centering
	\includegraphics[width=1\textwidth]{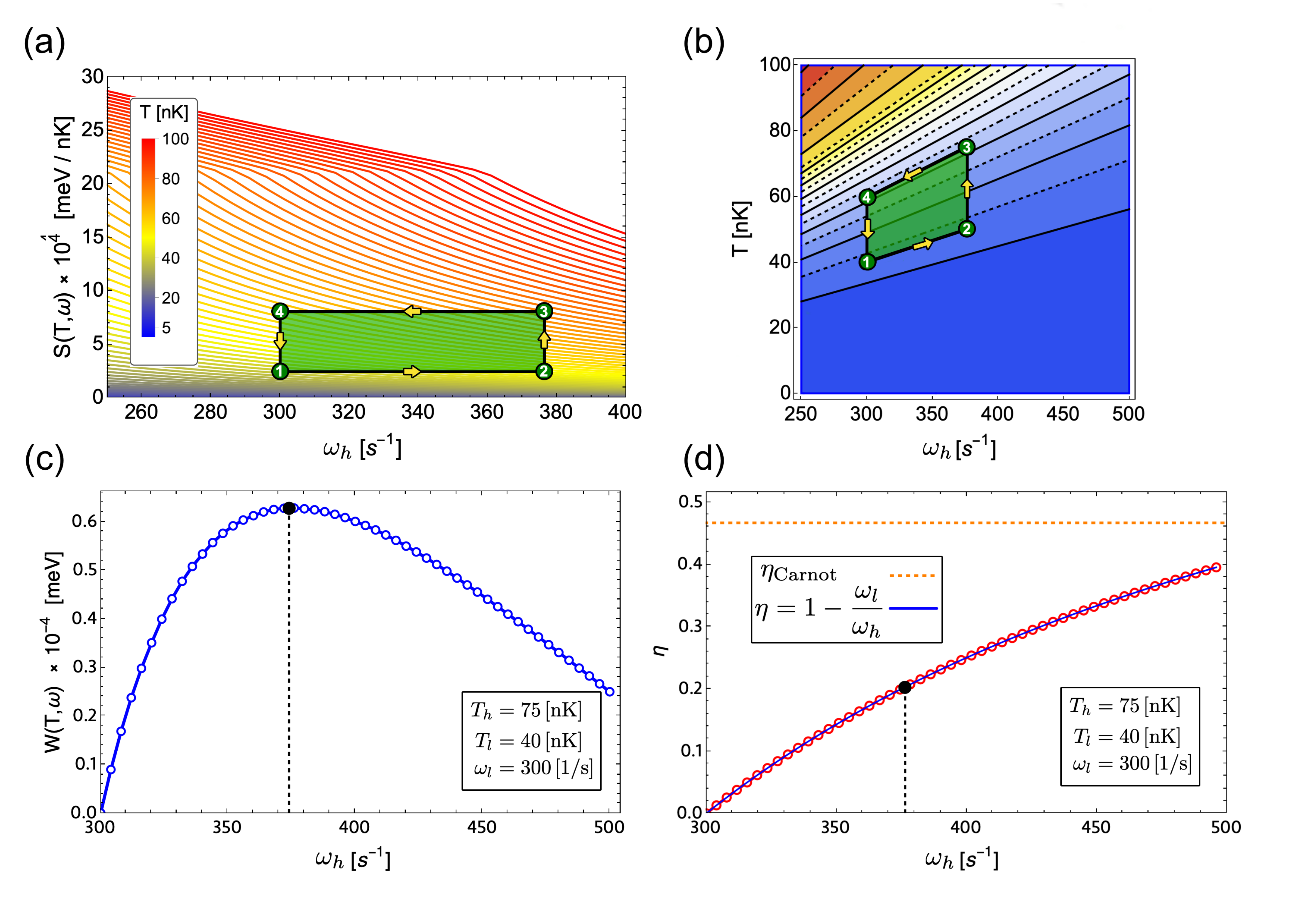}
	\caption{\label{fig:condensate} Results for a quasistatic cycle with 60,000 bosons operating fully in the condensate regime. (a) Representative cycle displayed over isothermal entropy curves. (b) $T$ versus $\omega_h$ plot of the cycle depicted in panel (a). (c) Total work as a function of $\omega_h$ for a range of cycles. (d) Efficiency as a function of $\omega_h$ for a range of cycles, with the analytical efficiency (solid blue line) and Carnot efficiency (dotted orange line) given for comparison. In panels (c) and (d) the work and efficiency values indicated with the black dots correspond to the specific cycle drawn in (a) and (b). The parameters are: $T_{l}= 40$ nK, $T_{h}=75$ nK, and $\omega_{l}=300 s^{-1}$.}
\end{figure}

\begin{figure}
	\centering
	\includegraphics[width=1\textwidth]{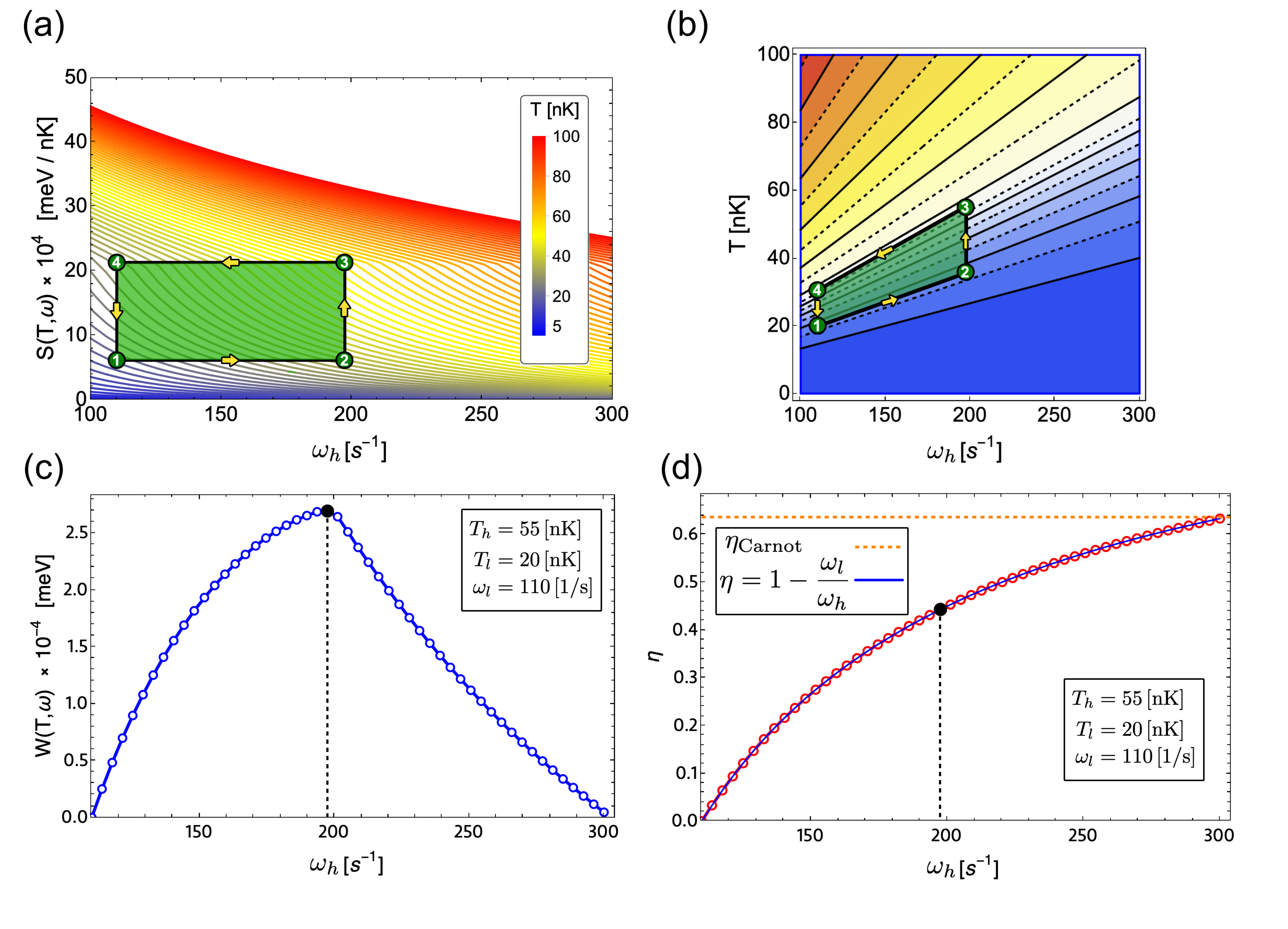}
	\caption{\label{fig:crosstrans} Results for a quasistatic cycle with 60,000 bosons driven across the condensate phase transition. (a) Representative cycle displayed over isothermal entropy curves. (b) $T$ versus $\omega_h$ plot of the cycle depicted in panel (a). (c) Total work as a function of $\omega_h$ for a range of cycles. (d) Efficiency as a function of $\omega_h$ for a range of cycles, with the analytical efficiency (solid blue line) and Carnot efficiency (dotted orange line) given for comparison. In panels (c) and (d) the work and efficiency values indicated with the black dots correspond to the specific cycle drawn in (a) and (b). The parameters are: $T_{l}= 20$ nK, $T_{h}=55$ nK, and $\omega_{l}=110 s^{-1}$.}
\end{figure}

Figures \ref{fig:noncondensate}, \ref{fig:condensate}, and \ref{fig:crosstrans} present numerical results for the quasistatic Otto cycle in three different operating zones, with the working medium in the non-condensate phase for the full cycle, with the working in the condensate phase for the full cycle, and with the working medium driven across the BEC phase transition during each isochoric stroke. In all the simulations, the parameters $T_{l}$, $T_{h}$ and $\omega_{l}$ are held fixed, while $\omega_{h}$ is varied. In each figure, panel (a) presents a representative cycle displayed over isothermal entropy curves as a function of $\omega$ for a temperature range between 5 nK and 350 nK. Panel (b) shows the same process presented in (a) but on a plot of $T$ versus $\omega_h$. Panel (c) shows the total work in units of meV over a wide range of possible cycles, with the black dot indicating the value of work obtained from the cycle highlighted in (a) and (b). Finally, panel (d) presents the efficiency as a function of $\omega$ for a range of cycles, where, as before, the efficiency value indicated with a black dot corresponds to the specific case drawn in (a) and (b). Note that the frequency and temperature values selected for our analysis are typical \cite{Grossmann1995} and comparable with those used in experimental demonstrations of Bose-Einstein condensation \cite{anderson1995observation, bradley1995evidence}.      

Figure \ref{fig:crosstrans} illustrates a case in which the working system transitions between the condensed and non-condensed phases during the isochoric strokes. In this case, the total work extraction is calculated from the combination of internal energy expressions given in Eq.~(\ref{internalenergy}). For stroke $1 \rightarrow 2$ the expression for $T \leq T_{c}$ correctly describes the internal energy of the working medium and for stroke $3 \rightarrow 4$ the expression for $T\geq T_{c}$ correctly describes the internal energy of the working medium. Consequently, the total work is given by, 
\begin{equation}
   \label{workcondensate-noncondesate}
W=-\frac{k_{B}^{4}\left(-1+ \kappa \right)\left(-\pi^{4}T_{l}^{4} + 90 T_{h}^{4} \kappa^{4} g_{4}(z_{h})\right)}{30 \kappa \omega_{1}^{3} \hbar^{3}},
\end{equation}
where we use $g_{4}(1)=\pi^{4}/90$. The intermediate temperatures, $T_{4}$ and $T_{2}$, were eliminated from Eq. \eqref{workcondensate-noncondesate} by applying the relationship between the temperatures and frequencies given in Eq. (\ref{temperaturerelationadiabatic}) that ensures the compression and expansion strokes remain isentropic.

The fugacity is obtained by solving Eq. \eqref{nthermodynamic} using a third-order approximation of the series form of $g_3(z)$,
\begin{equation}
    z + \frac{z^{2}}{2^{3}} + \frac{z^{3}}{3^{3}} = N \left(\frac{\hbar \omega}{k_{B} T}\right)^{3}.
\end{equation}
In order to verify that this third-order approximation is sufficiently accurate, we compare the analytical approximation to a numerical calculation of the fugacity in Appendix \hyperref[sec:AppendixB]{B}. 

It is important to note that only one Bose function appears in Eq.~(\ref{workcondensate-noncondesate}), despite the facts that the expressions for internal energy at both points 3 and 4 in the cycle are proportional to the Bose function, and that the temperatures and trap frequencies are different at each point. This is a consequence of the fugacity along path 3 $\rightarrow$ 4 in the cycle being obtained from Eq.~(\ref{nthermodynamic}) with a fixed number of particles, along with the isentropic condition given by Eq.~(\ref{temperaturerelationadiabatic}). The isentropic condition states that the ratio of frequency to temperature at the start and end of each adiabatic stroke must be equal. Since the fugacity is determined from this ratio, along with the number of particles, the fugacity must also remain fixed during the adiabatic strokes. Thus we have the additional condition that $z_{h}\equiv z_{h}(T_{h}, \omega_{2})= z_{4}(T_{4},\omega_{1})$.

Considering Eq. \eqref{critical} we see that, as long as the temperature of the hot reservoir remains fixed, increasing the trap frequency will eventually drive the system across the critical point. This transition results in a kink the total work extraction, due to the divergence in the first derivative of the internal energy that characterizes the BEC phase transition. This behavior can be clearly observed in Fig. \ref{fig:crosstrans}. By increasing $\omega_{h}$, we can move along the isotherm corresponding to $T_{h}= 55$ nK until we reach the frequency at which 55 nK is the critical temperature for the transition. We emphasize that Eq.~(\ref{workcondensate-noncondesate}) is only valid up to this transition point. By rearranging Eq. \eqref{critical} we can solve for this critical frequency, 
\begin{equation}
\label{criticalvalueofomega}
    \omega_{h}^{c}=\left(\frac{\zeta(3)}{N}\right)^{\frac{1}{3}}\left( \frac{k_{B} T_{c}}{\hbar}\right).
\end{equation}

Consequently, Eq.~(\ref{workcondensate-noncondesate}) is only valid up to $\kappa= \omega_l/\omega_{c}^{h}$. Rewriting Eq. \eqref{criticalvalueofomega} in terms of $\kappa$ and taking $T_c = T_h$ (which provides the limiting case that still ensures the phase transition occurs during the quasistatic heating stroke) we have,
\begin{equation}
    \kappa = 7.18\times 10^{-3} N^{1/3} \left(\frac{\omega_{l}}{T_{h}}\right).
\end{equation}
This equation indicates that there are two ways to decrease this critical value of $\kappa$ that ensures the engine is still operating in the transition regime (i.e. to avoid passing into the regime where the entire cycle takes place outside of the condensate phase): one is to decrease the number of particles, and the other is to decrease the ratio $\omega_{l}/T_h$. If we want to remain consistent with the assumption that the engine is operating in the thermodynamic limit, only the second option remains available. By decreasing the critical value of $\kappa$ we broaden the parameter space over which the transition engine can operate while still maintaining the condition that the compression stroke occurs in the condensate phase and that the expansion stroke occurs in the non-condensate phase.

Consider an example set of parameters, $\omega_{l}= 110$ s$^{-1}$, $T_{l} = 20$ nK, and $T_{h} = 55$ nK with $N= 60,000$ bosons. From Eq.~(\ref{criticalvalueofomega}) we obtain $\omega_{h}^{c}\sim 195.559$ s$^{-1}$, a value that is consistent with Fig. \ref{fig:crosstrans}(c). Beyond that value of $\omega_{h}$, to determine the total work extraction as shown in Fig. \ref{fig:crosstrans}(c) we must consider the case of a cycle that operates fully in the condensate regime.

Figure \ref{fig:condensate} presents such a case. As the working medium remains in the condensate phase for the full cycle, the fugacity ($z$) remains fixed at one. In this case, the total work of extraction is,
\begin{equation}
\label{workcondensate}
W=-\frac{k_{B}^{4}\pi^{4}\left(1+ \kappa \right)\left(-T_{l}^{4} + T_{h}^{4} \kappa^{4}\right)}{30 \kappa \omega_{1}^{3} \hbar^{3}},
\end{equation}
It is straightforward to maximize this expression for the work extraction in order to determine ideal the compression ratio. By taking the derivative of Eq. (\ref{workcondensate}) with respect to $\kappa$ and we obtain,
\begin{equation}
    T_{l}^{4} + T_{h}^{4} \left(3- 4\kappa^{*} \right) {\kappa^{*}}^{4} = 0.
\end{equation}
The example illustrated in Fig. \ref{fig:condensate} presents a cycle with parameters $T_{l} = 40$ nK, $T_{h} = 75$ nK and $\omega_{l} = 300$ s$^{-1}$. For these parameters, the work output is maximized at $\kappa^{*}= 0.7995$. The value of $\omega_{h}$ corresponding to this maximum is thus $\omega_{h} \sim 375. 232$ s$^{-1}$, which corresponds exactly with the peak observed in panel (c) of Fig. \ref{fig:condensate}.

It is important to note the fact that all our results for the engine operating only in the condensate regime are explicitly independent of the particle number, $N$. This is due to the fact that the chemical potential in the condensate phase is zero, resulting in $N$ no longer being a thermodynamic variable.

Finally, we will consider the case where the cycle is operated with a working medium that remains entirely in the non-condensate phase, as shown in Fig. \ref{fig:noncondensate}. In this case, the expression for the total work is, 
\begin{equation}
\label{worknoncondesate}
    W=- \frac{3k_{B}^{4}\left(-1+\kappa\right)\left(-T_{l} g_{4}(z_{l}) + T_{h}\kappa^{4} g_{4}(z_{h})\right)}{\kappa \omega_{1}^{3}\hbar^{3}},
\end{equation}
where the isentropic condition for the expansion and compression strokes leads to the following relations between the fugacities at each corner of the cycle, $z_{l}\equiv z_{l}(T_{l},\omega_{1})=z_{2}(T_{2}, \omega_{2})$ and $z_{h}\equiv z_{h}(T_{h}, \omega_{2})= z_{4}(T_{4},\omega_{1})$.

As in the case of the cycle fully in the condensate regime, we can use Eq. \eqref{worknoncondesate} to determine the compression ratio that maximizes the work output. Using the parameters of the example cycle shown in Fig. \ref{fig:noncondensate}, $T_{l}=120$ nK, $T_{h}=210$ nK, and $\omega_{l}=300$ s$^{-1}$, we obtain a value of $\kappa^{*}=0.755$. Using this value of $\kappa^*$, we see that the maximum occurs at $\omega_{h} \sim 398$ s$^{-1}$ consistent with the peak observed in panel (c) of Fig. \ref{fig:noncondensate}.


\subsection{Endoreversible results}

Let us first consider an endoreversible engine with a condensate working medium below the critical temperature. In this case, we can determine the efficiency by combining Eqs. \eqref{eq:eff}, \eqref{eq:Wcomp}, \eqref{eq:Wexp}, and \eqref{qinendo} with the top line of Eq. \eqref{internalenergy}. After making use of the isentropic conditions $T_A \omega_h = T_B \omega_l$ and $T_C \omega_l = T_D \omega_h$ the efficiency simplifies to the same result found for the quasistatic cycle,
\begin{equation}
    \label{eq:effBelow}
	\eta_{\mathrm{below}} = 1 - \kappa
\end{equation} 
where $\kappa = \omega_l/\omega_h$ is the compression ratio. 

We can repeat the same process for a non-condensate working medium above the critical temperature, now applying the expression for internal energy from the bottom line of Eq. \eqref{internalenergy}. Recalling that the fugacity remains constant during the isentropic strokes, we find that the dependence on the Bose function, $g_{\nu}(z)$, cancels out and the efficiency simplifies to,
\begin{equation}
    \label{eq:effAbove}
	\eta_{\mathrm{above}} = 1 - \kappa,
\end{equation}
identical to that of the condensate working medium. These results indicate that in both the quasistatic and endoreversible regimes Bose-Einstein condensation has no impact on engine efficiency. 

Next we consider the power output for a condensate working medium below the critical temperature. Combining Eq. \eqref{eq:P} with Eqs. \eqref{eq:Wcomp} and \eqref{eq:Wexp} yields a complicated expression in terms of the temperatures and frequencies at each corner of the cycle. Applying Eqs. \eqref{eq:isoheating} and \eqref{eq:isocooling} along with the isentropic conditions we can express the power entirely in terms of the experimentally controllable parameters, namely the hot and cold bath temperatures, the thermal conductivites, the stroke times, and the compression ratio,
\begin{equation}
\begin{split}
\label{eq:Pbelow}
P_{\mathrm{below}} = \frac{(k_B \pi)^4 (\kappa - 1)}{30\gamma\left(\tau_l + \tau_h\right)\left(e^{\alpha_l \tau_l+ \alpha_h \tau_h}-1\right)^4 \kappa^4 \omega_h^3 \hbar^3}&\Big\{\left[ \left(e^{\alpha_l \tau_l}-1\right)T_l+e^{\alpha_l \tau_l}\left(e^{\alpha_h \tau_h}-1\right)T_h \kappa\right]^4 \\
& - \left[T_h \kappa - e^{\alpha_h \tau_h}\left(\left(e^{\alpha_l \tau_l}-1\right)T_l+T_h \kappa\right)\right]^4 \Big\}.
\end{split}
\end{equation}
Following the same steps for a working medium above the critical temperature, we find the power to be,     
\begin{equation}
\begin{split}
\label{eq:Pabove}
	P_{\mathrm{above}} = & \frac{3 k_B^4 (\kappa - 1)}{\gamma\left(\tau_l+\tau_h\right)\left(e^{\alpha_l \tau_l+\alpha_h \tau_h}-1\right)^4 \kappa ^4 \omega_h^3 \hbar^3} \Big\{g_4(z_l) \left[\kappa  T_h-e^{\alpha_h \tau_h} \left(T_l \left(e^{\alpha_l \tau_l}-1\right)+\kappa T_h\right)\right]^4 \\
	& \qquad\qquad \qquad\qquad \qquad\qquad -g_4(z_h) \left[T_l \left(e^{\alpha_l \tau_l}-1\right)+\kappa  T_h e^{\alpha_l \tau_l} \left(e^{\alpha_h \tau_h}-1\right)\right]^4 \Big\},
\end{split}
\end{equation}
where $z_l$ and $z_h$ are the fugacities during the compression and expansion strokes, respectively.

It is well established that there is an inherent trade-off between efficiency and power. Efficiency is maximized in the limit of infinitely long, quasistatic strokes. However, in this limit the power output will vanish due to the dependence on stroke time in the denominator of Eq. \eqref{eq:P}. The maximum efficiency of the engine is bounded by the Carnot efficiency which, examining Eqs. \eqref{eq:effBelow} and \eqref{eq:effAbove}, we see is achieved when $\kappa = T_l/T_h$. Plugging this value of $\kappa$ into Eqs. \eqref{eq:Pbelow} and \eqref{eq:Pabove} we see that for both the condensate and non-condensate working mediums the power vanishes at Carnot efficiency.     

\begin{figure*}
	\centering
	\subfigure[]{
		\includegraphics[width=.31\textwidth]{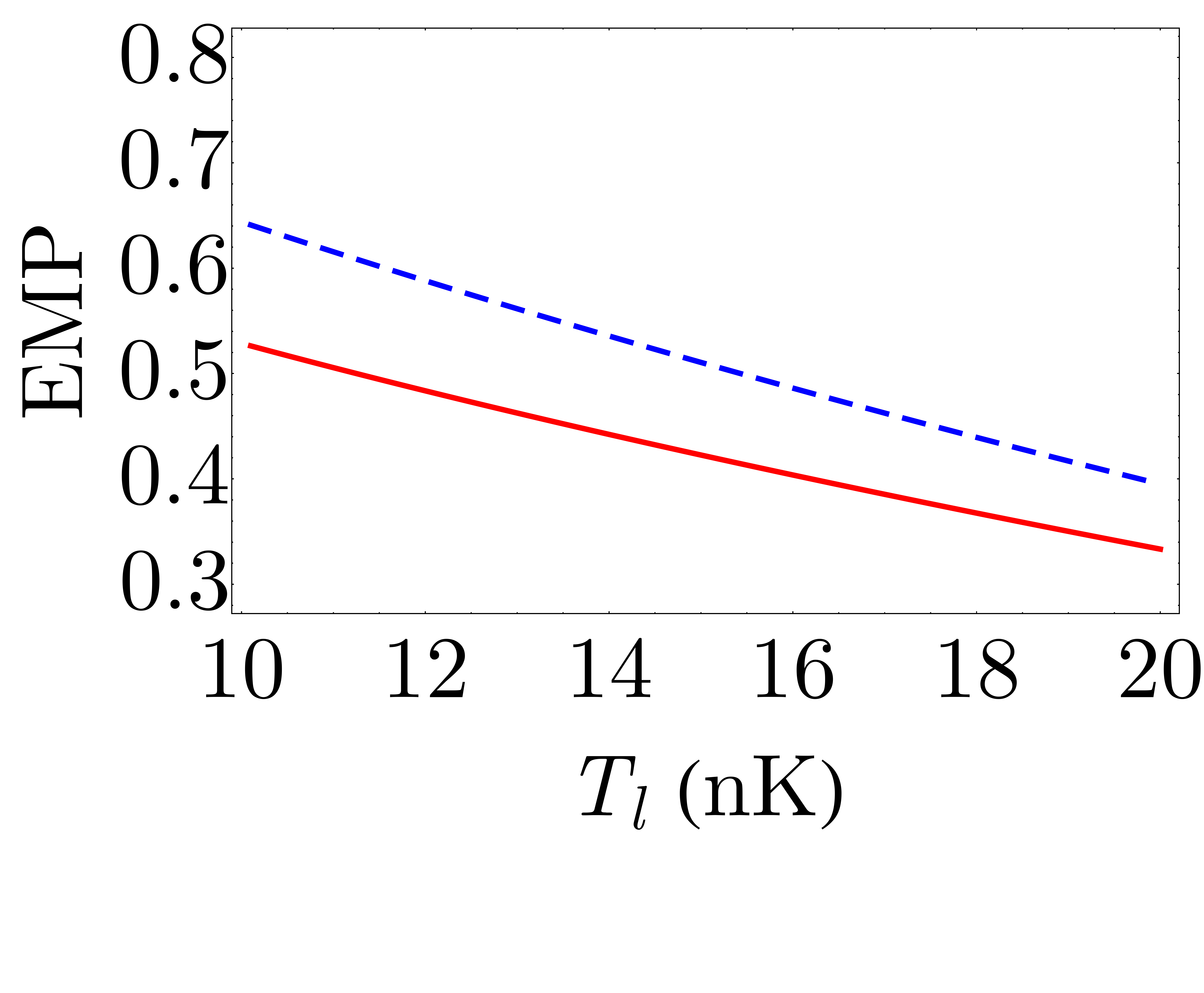}
	}
	\subfigure[]{
		\includegraphics[width=.31\textwidth]{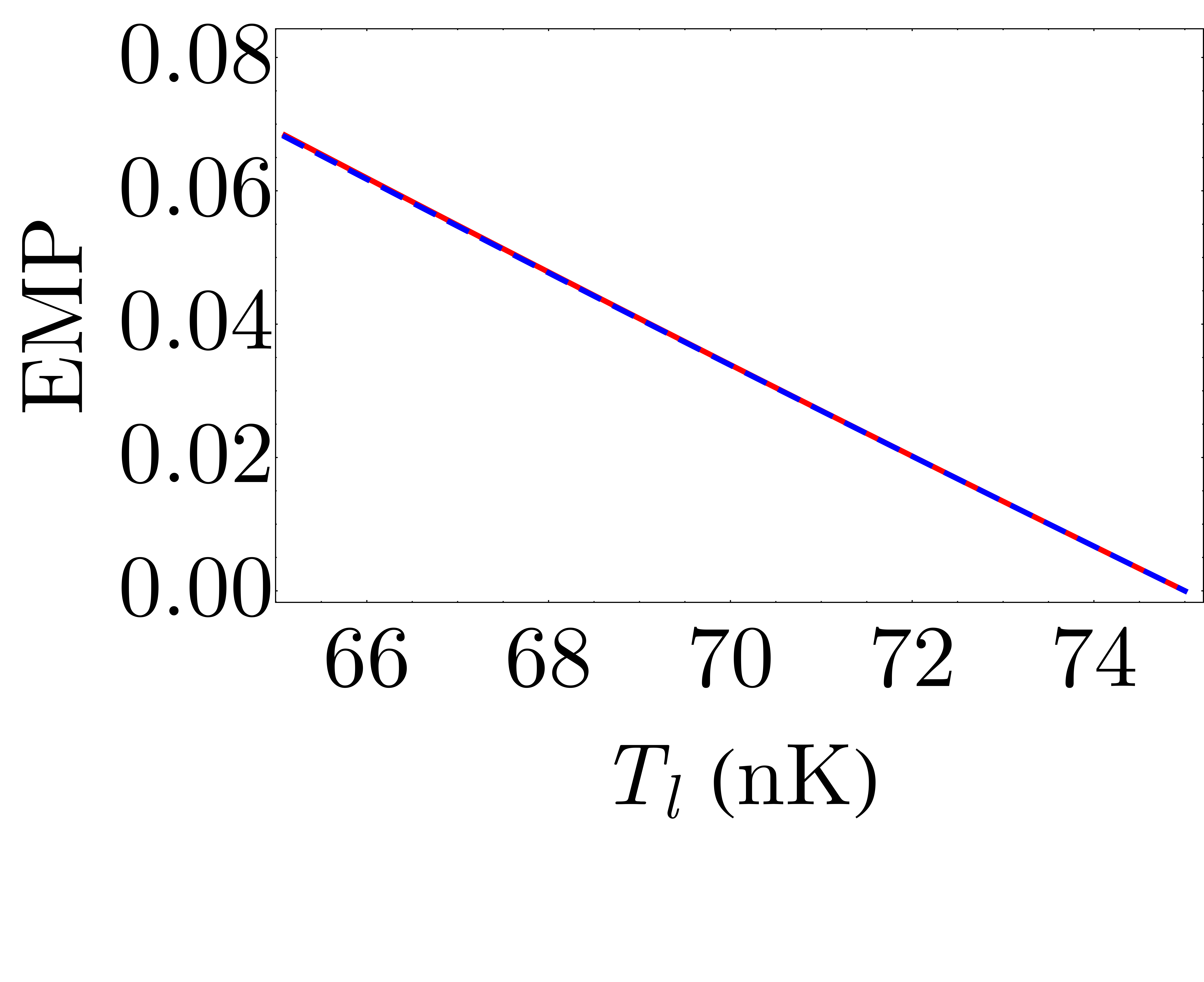}
	}
	\subfigure[]{
		\includegraphics[width=.31\textwidth]{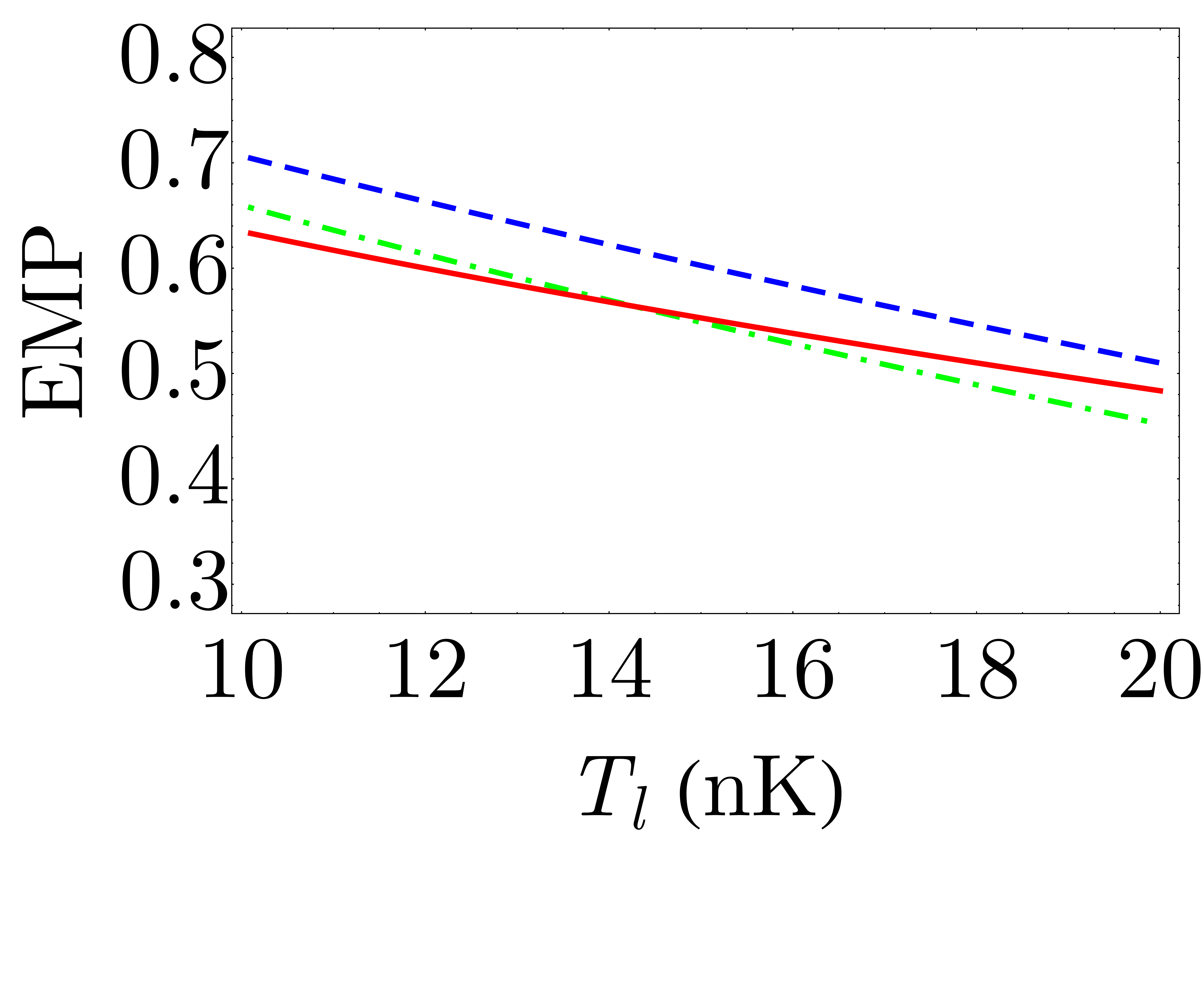}
	}
	\caption{\label{fig:EMPplot} (a) EMP as a function of the cold bath temperature for a BEC working medium (blue, dashed). Parameters are $\tau_c = \tau_h = 1$. We have used $T_h = 45$ nK, which ensures the temperature of the working medium always remains below the critical temperature. (b) EMP as a function of the cold bath temperature for a working medium of 60,000 bosons above the condensation threshold (blue, dashed). Parameters are $\tau_c = \tau_h = 1$. We have used $T_h = 75$ nK, which ensures the temperature of the working medium always remains above the critical temperature. (c) EMP as a function of the cold bath temperature for a working medium of 60,000 bosons driven across the BEC phase transition with stroke durations $\tau_c = \tau_h = 10$ (blue, dashed) and $\tau_c = \tau_h = 1$ (green, dot dashed). We have used $T_h = 75$ nK, which ensures that the compression stroke takes place below the critical temperature and the expansion stroke takes place above the critical temperature. In each plot the Curzon-Ahlborn efficiency (red, solid) is given for comparison. Other parameters are $\gamma = \alpha_c = \alpha_h = 1$.}
\end{figure*}

A figure of merit of more practical interest is the \textit{efficiency at maximum power} (EMP), which corresponds to maximizing the power output, and then determining the efficiency at that power \cite{Curzon1975}. As before, let us consider first the condensate working medium. Due to the cumbersome form of Eq. \eqref{eq:Pbelow} we maximize the power output numerically with respect to the compression ratio, $\kappa$. The EMP for a 60,000 boson condensate is shown in Fig. \ref{fig:EMPplot}a in comparison to the Curzon-Ahlborn efficiency. We see that the condensate EMP is significantly higher than the Curzon-Ahlborn efficiency. Noting that the Curzon-Ahlborn efficiency has been found to be the EMP for a classical harmonic Otto engine \cite{Deffner2018}, we see that the condensate behavior leads to a significant advantage in performance. 

We note that in Ref. \cite{Deffner2018} it was shown that a harmonic quantum Otto engine with a single-particle working medium displays an EMP that exceeds the Curzon-Ahlborn efficiency, as long as it is operating in the quantum regime defined by $\hbar \omega_h/k_B T_l \gg 1$. This is consistent with the results found here, as the parameters that ensure the working medium remains below the critical temperature throughout the cycle correspond to the deep quantum regime.

We next consider the EMP of a working medium of bosons above the critical temperature. Taking the ansatz that the high-temperature behavior should match the classical limit, we take the derivative of Eq. \eqref{eq:Pabove} with respect to $\kappa$ and then plug in our ansatz that $\kappa = \sqrt{T_l/T_h}$. Note that in order to express the compression and expansion stroke fugacities in closed form, we take the high temperature limit of small $z$ such that Eq. \eqref{eq:bosefunctionsum} can be well approximated by just the first term. We find that our ansatz works to maximize the power output in this case, demonstrating that the EMP above the critical temperature is equivalent to the Curzon-Ahlborn efficiency. This behavior is illustrated graphically in Fig. \ref{fig:EMPplot}b.  

Finally, we consider the case of a working medium of 60,000 bosons driven across the BEC phase transition, such that the compression stroke takes place while the working medium is below the critical temperature and the expansion stroke takes place while the working medium is above the critical temperature. The EMP for this transition engine is shown in Fig. \ref{fig:EMPplot}c for two different stroke durations. We see that when the duration of the isochoric strokes is short, the EMP of the bosonic medium is greater than the Curzon-Ahlborn efficiency at low values of the cold bath temperature but lower at higher values of the cold bath temperature. However, we see that if we increase the duration of the isochoric strokes, while the total power will be reduced, the EMP now exceeds the Curzon-Ahlborn efficiency across the whole temperature range. This indicates that the low value of the EMP observed at higher values of $T_l$ for short stroke times is a truly finite-time effect. We will explore the physical origins of this behavior in the next section.        

\section{Discussion}

\subsection{Work extraction from a BEC}

In the textbook formulation of a classical heat engine, work extraction occurs from the pressure exerted by the working medium against a movable piston during the expansion stroke of the cycle \cite{Callen}. However, this interpretation of work has clear issues when considering a fully condensed working medium, as particles in the zero-momentum state cannot exert any pressure \cite{Huang2009book}. This issue remains when considering the quantum formulation of work. For an isolated (unitarily evolving) quantum system, work is given by the change in internal energy \cite{Deffner2019book}. As the isentropic compression and expansion strokes of the quantum Otto cycle are performed while the working medium is isolated from the thermal environment, the total work extracted will be the sum of the changes in internal energy during both strokes. If all particles remain in the zero-momentum ground state across both strokes, the changes in internal energy will be the same, and the total work extracted from the engine will be zero.

Following this line of reasoning, we must think about how to interpret the results in Fig. \ref{fig:EMPplot}a. The engine is able to extract work while below the critical temperature, and does so at higher efficiency than a medium above the critical temperature. If the particles in the condensate contribute nothing to the work extracted from the engine, then this work extraction must come from the fraction of bosons that remain in an excited state outside of the condensate in the thermal cloud. The average number of particles in the thermal cloud can be found from \cite{Pathria2011},
\begin{equation}
\label{eq:Nexcited}
    N_T = \left(\frac{k_B T}{\hbar \omega} \right)^3 g_3(z).
\end{equation}
Utilizing the isentropic condition, along with the fact that the fugacity remains constant during the isentropic strokes, we see that the average number of particles in the thermal cloud must also remain fixed during the compression and expansion strokes.  

Let us consider a working medium below the critical temperature. In this regime $z=1$ and the internal energy can be described by the first line of Eq. \eqref{internalenergy}. Solving Eq. \eqref{eq:Nexcited} for $\omega$, we can rewrite the endoreversible work as,
\begin{equation}
\label{eq:NcompW}
    W_{\mathrm{comp}} = 3 k_B N^{\mathrm{comp}}_T \frac{\zeta (4)}{\zeta (3)}(T_2 - T_1),
\end{equation}
for the compression stroke and,
\begin{equation}
\label{eq:NexpW}
    W_{\mathrm{exp}} = 3 k_B N^{\mathrm{exp}}_T \frac{\zeta (4)}{\zeta (3)}(T_4 - T_3),
\end{equation}
for the expansion stroke. Using Eqs. \eqref{eq:NcompW} and \eqref{eq:NexpW} as well as Eqs. \eqref{eq:isoheating} and \eqref{eq:isocooling} and the isentropic conditions, we can express the endoreversible power entirely in terms of the experimental controllable parameters along with $N^{\mathrm{comp}}_T$ and $N^{\mathrm{exp}}_T$,
\begin{equation}
\label{eq:Npower}
    P = \frac{\pi^4 (1-\kappa)\hbar \omega_2}{30\gamma \left(\tau_c + \tau_h\right)\zeta(3)^{4/3}}\left[\left(N^{\mathrm{exp}}_T\right)^{4/3} - \left(N^{\mathrm{comp}}_T\right)^{4/3} \right].
\end{equation}
From this expression it is clear that the power depends directly on the number of particles in the thermal cloud during the compression and expansion strokes. Furthermore, we see that if all particles reside in the condensate, such that $N^{\mathrm{exp}}_T = N^{\mathrm{comp}}_T = 0$, the power output vanishes, confirming our supposition that the work extracted from the engine comes entirely from the bosons that remain in the thermal cloud. 

From Eq. \eqref{eq:Npower} we see that the power is maximized when $N^{\mathrm{exp}}_T$ is as large as possible and $N^{\mathrm{comp}}_T$ is as small as possible. Examining Eq. \eqref{critical} we see that $N_T$ achieves its maximum value of $N$ when $T = T_c$, and vanishes as $T$ approaches zero. Thus in order to maximize the power output from the engine, we want the expansion stroke to be as far below the critical temperature as possible, and the compression stroke to occur as close to the critical temperature as possible.

This provides a straightforward physical interpretation of the enhanced performance we see for the BEC working medium in comparison to a working medium above the critical temperature. As particles in the zero-momentum state cannot exert any pressure, the compressibility of the BEC phase diverges. With this being the case, just as no work can be extracted from the bosons in the BEC during the expansion stroke, no work is needed to compress them during the compression stroke. However, after the isochoric heating stroke a fraction of the particles that were compressed ``for free" in the condensate will have been excited into the thermal cloud, allowing them to do work during the expansion process.

Let us now consider how this behavior leads to the EMPs seen in Figs. \ref{fig:EMPplot}a, \ref{fig:EMPplot}b, and \ref{fig:EMPplot}c. In order to maximize the power output the work done on the medium during the compression stroke should be minimized, which occurs when $T_1 \approx T_2$, and the work by the medium during the expansion stroke should be maximized, which occurs when $T_3 \gg T_4$. Recalling the isentropic condition in Eq. \eqref{temperaturerelationadiabatic} we see that when $\kappa \approx 1$ the work cost of compression will be minimized, but when $\kappa \ll 1$ the work gained on expansion will be maximized. The maximal power output occurs at the value $\kappa^*$ that best balances this trade-off. 

Since the efficiency is always given by $\eta = 1-\kappa$, regardless of the phase of the working medium, the fact that the BEC medium exceeds the Curzon-Ahlborn efficiency means $\kappa^*$ is always less than $(T_l/T_h)^{1/2}$ when below the critical temperature. As the BEC medium can be compressed at a lower work cost, $\kappa^*$ can shift to a lower value, increasing the work gained on expansion. The same interpretation can be applied for the transition engine when $T_l$ is significantly below the critical temperature. 

However, we see in Fig. \ref{fig:EMPplot}c that at larger values of $T_l$, when the temperature of the cold bath is closer to the critical temperature, the EMP falls below Curzon-Ahlborn. Using Eq. \eqref{critical} we can express the critical temperature for the cooling stroke as,
\begin{equation}
    \label{eq:coldCrit}
    T_c^{\mathrm{cool}} = \frac{\hbar \kappa \omega_2}{k_B}\left(\frac{N}{\zeta(3)}\right)^{1/3}. 
\end{equation}
From Eq. \eqref{eq:coldCrit} we see that the critical temperature also depends on $\kappa$. Thus having a larger value of $\kappa$ raises the critical temperature for the cooling stroke, resulting in a cycle that operates with a larger percentage in the BEC regime and further reducing the compression work cost. This leads to larger values of $\kappa^*$ being favorable for maximum power output, leading to reduced EMP. However, as $T_l$ gets colder and colder, this increases the percentage of the cycle in the BEC regime faster than increasing $\kappa$ would, and thus smaller values of $\kappa^*$ that maximize the work gained on expansion become preferable.

The fact that increasing the cycle times leads to the transition engine EMP always exceeding the Curzon-Ahlborn efficiency favors this interpretation. Increasing the stroke times has a similar effect to lowering $T_l$ (as both lead to a decrease in $T_4$). Like decreasing $T_l$, increasing the stroke times moves a larger percentage of the cycle into the BEC regime faster than raising the critical temperature by increasing $\kappa$. Thus for the case of long stroke times smaller values of $\kappa^*$ become preferable again. 

\subsection{Experimental considerations}

Typically, BECs are created by cooling a trapped atomic gas using evaporative cooling, laser cooling, or a combination of the two \cite{anderson1995observation, davis1995bose, bradley1995evidence, Ketterle1996, Barrett2001}. While laser cooling is capable of achieving temperatures of only a few nanokelvin, it is most successful in low density systems \cite{Ketterle1996}. For systems operating in the thermodynamic limit with a  large number of particles, evaporative cooling provides a more realistic approach for achieving condensation.

Evaporative cooling provides an additional complication not explicitly considered in our analysis, namely that the total number of particles is no longer fixed. During the cooling strokes, the evaporative cooling process will result in a decrease in particle number, precluding the possibility of a completely closed thermodynamic cycle. However, this does not mean our analysis lacks applicability. 

The simplest scenario under which our results remain valid is if the number of particles is sufficiently large, such that the fraction lost during each cooling stroke remains effectively zero. Furthermore, for an engine operating below the critical temperature, the total number of particles is not a thermodynamic variable, with the number of particles in the thermal cloud being determined solely by the temperature and frequency. Thus for the case of an engine operating fully in the condensate regime, as long as the number of particles lost to evaporative cooling remains low enough that the assumption of the thermodynamic limit is still valid, our results will remain applicable.         

\subsection{Concluding remarks}

In this work we have examined both the quasistatic and endoreversible performance of an Otto cycle with a working medium of a harmonically confined Bose gas. We have shown that when the cycle is operated above the critical temperature, the efficiency at maximum power is equivalent to the Curzon-Ahlborn efficiency. However, when the cycle is operated below the critical temperature the efficiency at maximum power can significantly exceed the Curzon-Ahlborn efficiency. We have demonstrated that the power output of such a cycle is optimal when the number of particles in the condensate is maximized during the compression stroke and the number of particles in the thermal cloud is maximized during the expansion stroke. This enhanced power output is fundamentally an effect of the indistinguishable nature of quantum particles, arising from the fact that the particles in the zero-momentum state can be compressed at no work cost.

Bose-Einstein condensates show much potential for the development of ultra-high precision sensors \cite{Aguilera2014} and for applications in quantum information processing \cite{Byrnes2012}. However, in order to optimally implement BEC-based devices we must first understand their thermodynamics in a device-oriented context. Heat engines provide just such a framework. Here we have shown that the unique properties of the BEC phase can be leveraged to enhance heat engine performance. As the BEC phase is a fundamentally quantum state of matter, this is a demonstration of a thermodynamic ``quantum advantage." 

It has also been demonstrated that the phenomena of Bose condensation can occur outside the realm of ultra-cold atomic gasses, such as in magnons, quasiparticle spin excitations in magnetic systems \cite{Giamarchi2008}. Magnon condensates have the distinct advantage of surviving at much higher temperatures, even up to room temperature \cite{Demokritov2006}. Such systems may provide an ideal platform for experimental implementations of a BEC engine. Strongly coupled photon-matter excitations, known as polaritons \cite{Hopfield1958, Carusotto2013}, can also exhibit condensation. Polaritons are a fundamentally nonequilibrium system \cite{Carusotto2013}, however, and their nonequilibrium nature would have to be carefully accounted for in any treatment of their thermodynamics.  

In this paper we have considered a BEC operating in the quasistatic and endoreveresible regimes. In the endoreversible regime, the protocol by which the frequency is varied during the compression and expansion strokes is irrelevant, as long as the condition of local equilibrium is maintained. Extending this work to the fully nonequilibrium regime would allow for an exploration of non-equilibrium, finite-time effects, such as the impact of specific ramp protocols on the engine performance. In Ref. \cite{Li2018} it was shown that work can be extracted from a BEC in the nonequilibrium regime by varying the nonlinear interaction strength of the BEC through the use of Feshbach resonances. Extending our analysis to the nonequilibrium regime would introduce the possibility of a cycle that can leverage both variations in the nonlinearity strength and external trapping potential. In principle, the work extracted from a BEC engine could also be employed to refrigerate another coupled gas. This might provide an avenue towards an effective means of cooling ultra-cold atomic vapors beyond the evaporative or laser cooling paradigms. We leave an exploration of these questions and more as potential topics for future work.   

\section*{Acknowledgments}

This material is based upon work supported by the U.S. Department of Energy, Office of Science, Office of Workforce Development for Teachers and Scientists, Office of Science Graduate Student Research (SCGSR) program. The SCGSR program is administered by the Oak Ridge Institute for Science and Education for the DOE under contract number DE‐SC0014664. S.D. acknowledge support from the U.S. National Science Foundation under Grant No. DMR-2010127. F.J.P. acknowledges support from ANID Fondecyt, Iniciaci\'on en Investigaci\'on 2020 grant No. 11200032, and the financial support of USM-DGIIE. P.V. acknowledges support from ANID Fondecyt grant No. 1210312 and to ANID PIA/Basal grant No.  AFB18000. O. N. acknowledges support from to ANID PIA/Basal grant No.  AFB18000. G.D.C. acknowledges support by the UK EPSRC EP/S02994X/1.

\appendix

\section*{Appendix A: Isentropic condition}
\renewcommand{\theequation}{A.\arabic{equation}}
\setcounter{equation}{0} 
\label{sec:AppendixA}

In this appendix we will derive the relationship between the frequency and temperature that maintains the isentropic condition for the compression and expansion strokes. From Eq.~(\ref{entropy}) we know that for $T\geq T_{c}$ the entropy is given by, 
\begin{equation}
\label{eq:entropyAbove}
    S(T, \omega) = k_{B}\left(\frac{k_{B} T}{\hbar \omega}\right)^{3}\zeta (3) \left[\frac{4}{N}\left(\frac{k_{B} T}{\hbar \omega}\right)^{3}g_{4}(z)-\ln (z)\right].
\end{equation}
Noting that the entropy for $T < T_c$ is simply a special case of Eq. \eqref{eq:entropyAbove} with $z = 1$, if we can determine a relationship that maintains $dS = 0$ for arbitrary $z$ we know that it will hold both above and below the critical temperature. As $S = S(T, \omega)$ we can express the isentropic condition as,
\begin{equation}
\label{eq:isenCond}
    dS = \left(\frac{\partial S}{\partial T}\right)_{\omega} dT + \left(\frac{\partial S}{\partial \omega}\right)_T d\omega = 0.
\end{equation}
Re-arranging Eq.~(\ref{eq:isenCond}) we arrive at Eq.~(\ref{differential}),
\begin{equation}
    \frac{d\omega}{dT}=-\frac{\left(\frac{\partial S}{\partial T}\right)_{\omega}}{\left(\frac{\partial S}{\partial \omega}\right)_{T}}.
\end{equation} 
Taking the appropriate derivatives of Eq.~(\ref{eq:entropyAbove}) we arrive at the cumbersome expression, 
\begin{equation}
\label{totalderivative}
     -\frac{\left(\frac{\partial S}{\partial T}\right)_{\omega}}{\left(\frac{\partial S}{\partial \omega}\right)_{T}}=\frac{\omega}{T}\left(\frac{\left[24 k_{B}^{3} T^{3} g_{4}(z) z - N T \hbar^{3}\omega^{3}\frac{\partial z}{\partial T} + z \left(4 k_{B}^{3} T^{3} \frac{\partial g_{4}(z)}{\partial z} \frac{\partial z}{\partial T} - 3N\hbar^{3} \omega^{3}\ln \left( z \right)\right)\right]}{ \left[24 k_{B}^{3} T^{3} g_{4}(z) z  + N\hbar^{3}\omega^{4}\frac{\partial z}{\partial \omega} - z \left(4k_{B}T^{3}\omega \frac{\partial g_{4}(z)}{\partial z}\frac{\partial z}{\partial \omega} + 3 N\hbar^{3}\omega^{3}\ln\left(z\right)\right)\right] }\right).
\end{equation}

Let us consider the ansatz that the temperature-frequency relationship that maintains the isentropic condition is identical to the relationship found for a single-particle harmonic Otto engine in Ref. \cite{Deffner2018}, that is $T_A \omega_h = T_B \omega_l$ and $T_C \omega_l = T_D \omega_h$. If this is the case, then we need to show that the right hand side of Eq.~(\ref{totalderivative}) simplifies to $\omega/T$. Thus our ansatz is verified if we can show the term in the large $\left( \right)$ in the Eq.~(\ref{totalderivative}) is equal to one. Examining Eq.~(\ref{totalderivative}) we see that there are two conditions under which this will be true. Either    
\begin{equation}
\label{firstcondition}
\left(N\hbar^{3}\omega^{3}-4k_{B}^{3}T^{3}z\frac{\partial g_{4}(z)}{\partial z}\right) = 0,
\end{equation}
or
\begin{equation}
\label{secondcondition}
    \left(T\frac{\partial z}{\partial T} + \omega \frac{\partial z}{\partial \omega}\right) = 0.
\end{equation}

Let us consider the first condition, presented in Eq.~(\ref{firstcondition}). This condition will be satisfied if, 
\begin{equation}
\label{eq:CondOneEq}
    z\frac{\partial g_{4}(z)}{\partial z} = \frac{N}{4}\left(\frac{\hbar \omega}{k_B T}\right)^3.
\end{equation}
The function $g_{\nu}(z)$ obeys the recurrence relation \cite{Pathria2011},
\begin{equation}
    g_{\nu - 1} (z)=\frac{\partial}{\partial \ln(z)} g_{\nu}(z),
\end{equation}
which we can use to express Eq.~(\ref{eq:CondOneEq}) as,
\begin{equation}
   g_3(z) = \frac{N}{4}\left(\frac{\hbar \omega}{k_B T}\right)^3.
\end{equation}
However, using Eq.~(\ref{nthermodynamic}) we know that,  
\begin{equation}
   g_3(z) = N\left(\frac{\hbar \omega}{k_B T}\right)^3.
\end{equation}
Thus, the first condition simplifies to $1 = 1/4$, which can never be satisfied.

Next we examine the second condition, presented in Eq.~(\ref{secondcondition}). This can be rewritten,
\begin{equation}
\label{todemonstrate}
   T \frac{\partial z}{\partial T} = - \omega \frac{\partial z}{\partial \omega}.
\end{equation}
By applying the chain rule, along with Eq.~(\ref{nthermodynamic}), we arrive at,
\begin{equation}
    \frac{\partial g_{3}(z)}{\partial \omega} = \frac{\partial g_{3}(z)}{\partial z} \frac{\partial z}{\partial \omega} =  3N \left(\frac{\hbar}{k_{B} T}\right)^{3} \omega^{2}.
\end{equation}
Thus,
\begin{equation}
\label{firstfrac}
    \frac{\partial z}{\partial \omega} = 3 N \left(\frac{\hbar \omega}{k_{B} T}\right)^{3} \frac{1}{\omega} \left[\frac{\partial g_{3}(z)}{\partial z}\right]^{-1}.
\end{equation}
Similarly, we can again apply Eq.~(\ref{nthermodynamic}) to find, 
\begin{equation}
    \frac{\partial g_{3}(z)}{\partial T}=\frac{\partial g_{3}(z)}{\partial z}\frac{\partial z}{\partial T}=- 3N \left(\frac{\hbar \omega}{k_{B}}\right)^{3}\frac{1}{T^{4}}.
\end{equation}
Therefore,
\begin{equation}
\label{secondfrac}
    \frac{\partial z}{\partial T}=- 3N \left(\frac{\hbar \omega}{k_{B} T}\right)\frac{1}{T} \left[\frac{\partial g_{3}(z)}{\partial z}\right]^{-1}.
\end{equation}
Finally, comparing Eq.~(\ref{secondfrac}) and Eq.~(\ref{firstfrac}) we confirm that Eq. (\ref{todemonstrate}) is indeed true and holds for arbitray $z$. Consequently, the linear relationship holds for all temperatures with no approximations.

\section*{Appendix B: Numerical fugacity calculation}
\renewcommand{\theequation}{B.\arabic{equation}}
\setcounter{equation}{0} 
\renewcommand{\thefigure}{B\arabic{figure}}
\setcounter{figure}{0}
\label{sec:AppendixB}

\begin{figure}[h]
	\centering
	\includegraphics[width=.55\textwidth]{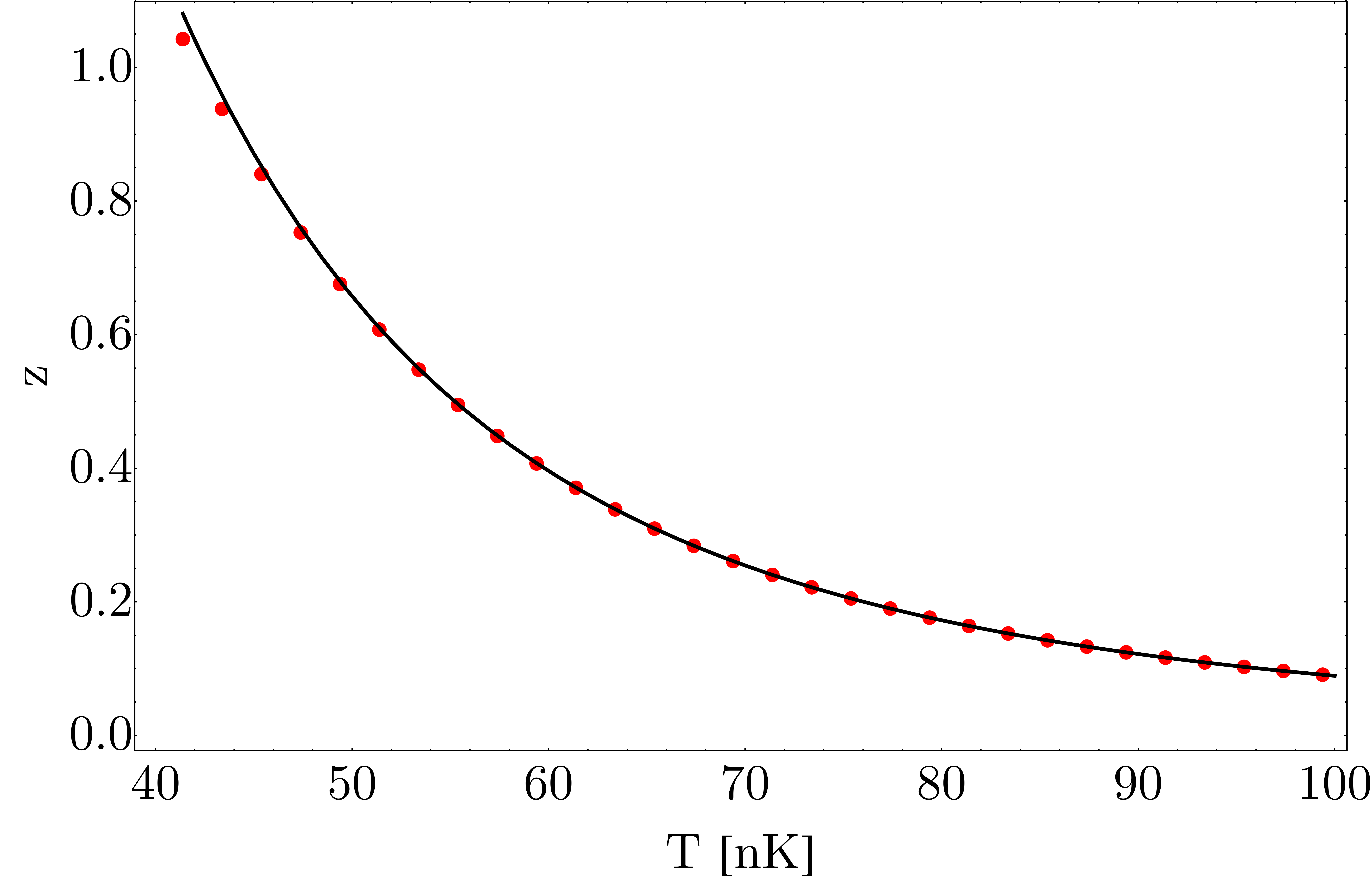}
	\caption{\label{fig:zCompPlot} Comparison between numerical calculation (red dots) and analytical third-order approximation (black line) for the fugacity of a 60,000 boson system. Trap frequency was taken to be $\omega = 150 s^{-1}$.}
\end{figure}

In this appendix we verify the fugacity found using the the third-order analytical approximation using a direct numerical calculation. In the numerical case, the fugacity is found by solving for the chemical potential at which the average particle number, determined by summing over the first 10,000 state occupations of the Bose-Einstein distribution, is equivalent to the total particle number. The temperature range and trap frequency parameters were chosen to be comparable to those used for the example cycle in Fig. \ref{fig:crosstrans}. We see in Fig. \ref{fig:zCompPlot} that the analytical approximation and numerical results show very good agreement, with the third-order approximation deviating only slightly when very close to the critical temperature.     

\section*{References}

\bibliographystyle{iopart-num}
\bibliography{BECRefs}

\end{document}